\documentclass[final]{svjour2}
\usepackage{graphicx}
\usepackage{rotating}
\usepackage{amssymb}
\usepackage{hyperref}
\usepackage{mathptmx}
\usepackage[numbers]{natbib}
\makeatletter
\journalname{Journal of Low Temperature Physics}

\bibpunct{}{}{,}{s}{}{,}

\begin{document}

\newcommand{\hdblarrow}{H\makebox[0.9ex][l]{$\downdownarrows$}-}
\title{The NIKA2 instrument at 30-m IRAM telescope: performance and results}

\author{
A.~Catalano$^{9,7}$ (\email{catalano@lpsc.in2p3.fr} )
\and R.~Adam$^{1}$
\and  P.A.R.~Ade $^{3}$
\and  P.~Andr\'e $^{4}$
\and  H.~Aussel $^{4}$
\and  A.~Beelen $^{6}$
\and  A.~Beno\^it $^{7}$
\and  A.~Bideaud $^{7}$
\and  N.~Billot $^{8}$
\and  O.~Bourrion $^{9}$
\and  M.~Calvo $^{7}$
\and  B.~Comis $^{9}$
\and  M. De Petris $^{11}$ 
\and  F.-X.~D\'esert $^{5}$
\and  S.~Doyle $^{3}$
\and  E.F.C.~Driessen $^{10}$
\and  J.~Goupy $^{7}$
\and  C.~Kramer $^{8}$
\and  G.~Lagache $^{6}$
\and  S.~Leclercq $^{10}$
\and  J.-F.~Lestrade $^{12}$
\and  J.F.~Mac\'ias-P\'erez $^{9}$
\and  P.~Mauskopf $^{13}$
\and  F.~Mayet $^{9}$
\and  A.~Monfardini $^{7}$ 
\and  E.~Pascale $^{3}$
\and  L.~Perotto $^{9}$
\and  G.~Pisano $^{3}$
\and  N.~Ponthieu $^{5}$
\and  V.~Rev\'eret $^{4}$
\and A.~Ritacco $^{8}$
\and  C.~Romero $^{10}$
\and  H.~Roussel $^{14}$
\and  F.~Ruppin $^{9}$
\and  K.~Schuster $^{10}$
\and  A.~Sievers $^{8}$
\and  S.~Triqueneaux $^{7}$
\and  C.~Tucker $^{3}$
\and  R.~Zylka $^{10}$
\and E.~Barria $^{7}$
\and G.~Bres $^{7}$
\and P.~Camus $^{7}$ 
\and P.~Chanthib $^{7}$ 
\and G.~Donnier-Valentin $^{7}$ 
\and O.~Exshaw $^{7}$ 
\and G.~Garde $^{7}$ 
\and A.~Gerardin $^{7}$
\and J.-P.~Leggeri $^{7}$
\and F.~Levy-Bertrand $^{7}$
\and C.~Guttin $^{7}$ 
\and C.~Hoarau $^{7}$ 
\and M.~Grollier $^{7}$ 
\and J.-L.~Mocellin $^{7}$ 
\and G.~Pont $^{7}$ 
\and H.~Rodenas $^{7}$ 
\and O.~Tissot $^{7}$
\and G.~Galvez $^{8}$
\and D.~John $^{8}$ 
\and H.~Ungerechts $^{8}$ 
\and S.~Sanchez $^{8}$ 
\and P.~Mellado $^{8}$ 
\and M.~Munoz $^{8}$ 
\and F.~Pierfederici $^{8}$ 
\and J.~Penalver $^{8}$ 
\and S.~Navarro $^{8}$
\and G.~Bosson $^{9}$
\and J.-L.~Bouly $^{9}$
\and J.~Bouvier $^{9}$
\and C.~Geraci $^{9}$
\and C.~Li $^{9}$
\and J.~Menu $^{9}$
\and N.~Ponchant $^{9}$
\and S.~Roni $^{9}$
\and S.~Roudier $^{9}$
\and J.P.~Scordillis $^{9}$
\and D.~Tourres $^{9}$
\and C.~Vescovi $^{9}$
\and A.~Barbier $^{10}$
\and D.~Billon-Pierron $^{10}$ 
\and A.~Adane $^{2}$ 
\and  A.~Andrianasolo $^{5}$  %
\and  A.~Bracco $^{4}$  %
\and  G.~Coiffard $^{10}$ %
\and  R.~Evans $^{3}$  %
\and  A.~Maury $^{4}$  %
\and  A.~Rigby $^{3}$  %
}

%
%
%
%

\institute{
1 : Laboratoire Lagrange, Universit\'e et Observatoire de la C\^ote d'Azur, CNRS, Blvd de l'Observatoire, F-06304 Nice, France\label{OCA} \\
2 : University of Sciences and Technology Houari Boumediene (U.S.T.H.B.), BP 32 El Alia, Bab Ezzouar 16111, Algiers, Algeria \label{USTHB} \\
3 : Astronomy Instrumentation Group, University of Cardiff, United Kindgom\label{Cardiff} \\
4 : Laboratoire AIM, CEA/IRFU, CNRS/INSU, Universit\'e Paris Diderot, CEA-Saclay, 91191 Gif-Sur-Yvette, France\label{CEA} \\
5 : Institut de Plan\'etologie et Astrophysique de Grenoble (IPAG), UGA \& CNRS, BP 53, F-38041 Grenoble, France\label{IPAG} \\
6 : Aix Marseille Universit\'e, CNRS, LAM (Laboratoire d'Astrophysique de Marseille), F-13388 Marseille, France \label{LAM} \\
7 : Institut N\'eel, CNRS and Universit\'e Grenoble Alpes (UGA), 25 av. des Martyrs, F-38042 Grenoble, France \label{Neel} \\
8 : Institut de RadioAstronomie Millim\'etrique (IRAM), Granada, Spain \label{IRAME} \\
9 : Laboratoire de Physique Subatomique et de Cosmologie, Universit\'e Grenoble Alpes, CNRS, 53, av. des Martyrs, Grenoble, France\label{LPSC} \\
10 : Institut de RadioAstronomie Millim\'etrique (IRAM), Grenoble, France \label{IRAMF} \\
11 : Dipartimento di Fisica, Universit\`a di Roma La Sapienza, Piazzale Aldo Moro 5, I-00185 Roma, Italy \label{Roma} \\ 
12 : LERMA, CNRS, Observatoire de Paris, 61 avenue de l'Observatoire, Paris, France \label{LERMA} \\
13 : School of Earth and Space Exploration and Department of Physics, Arizona State University, Tempe, AZ 85287 \label{Arizona} \\
14 : Institut d'Astrophysique de Paris, CNRS (UMR7095), 98 bis boulevard Arago, F-75014 Paris, France \label{IAP} 
 }

\date{30.06.2017}

\maketitle

\begin{abstract}

The New IRAM KID Arrays~2 (NIKA2) consortium has just finished installing and commissioning a millimetre camera on the IRAM 30 m telescope. It is a dual-band camera operating with three frequency multiplexed kilo-pixels arrays of Lumped Element Kinetic Inductance Detectors (LEKID) cooled at 150~mK, designed to observe the intensity and polarisation of the sky at 260 and 150~GHz (1.15 and 2~mm). NIKA2 is today an IRAM resident instrument for millimetre astronomy, such as Intra Cluster Medium from intermediate to distant clusters and so for the follow-up of Planck satellite detected clusters, high redshift sources and quasars, early stages of star formation and nearby galaxies emission. We present an overview of the instrument performance as it has been evaluated at the end of the commissioning phase. 

\keywords{Millimetre Astrophysics, Detectors LEKID}

\end{abstract}

\section{Introduction}\label{intro}

New frontiers in millimetre astronomy require high sensitivity and high resolution instruments. These goals demand the development of large-format instruments with arrays of detectors. The technological solution that we have chosen uses Lumped Element Kinetic Inductance Detectors (LEKID). It allows for a large multiplexing factor frequency domain readout and an accessible manufacturing. This technological solution has been selected for the NIKA project which represents the first demonstration of LEKID performance using kilo-pixel arrays for scientific observations at millimetre wavelength\cite{catalano_nika2014,adam2014}. The camera has been permanently installed at the IRAM 30~m telescope in October 2015, and it is available for the general community since summer semester 2017.


The NIKA2 consortium is responsible for the design, the construction, and the commissioning of the instrument. In addition, NIKA2 consortium takes the pledge to provide technical support for the 10 years of foreseen life time of the instrument. This contribution is rewarded with 1300 hours of guaranteed time distributed in five large programs over 4 years :

 \begin{itemize}

\item  Clusters of galaxies via the Sunyaev Zel'dovitch effect

\item  Deep surveys (mapping large areas at a depth close to the confusion limit)

\item  Mapping the insterstellar medium

\item  Nearby Galaxies

\item  Polarization measurements of Galactic regions

 \end{itemize}

In this paper we report the performance obtained at the end of the commissioning  in terms of instrumental noise equivalent flux density (NEFD) calculated on point-like sources. This work has been extensively discussed in a recent NIKA 2 paper \cite{nika2}

\section{NIKA2 instrument and the whole calibration process}\label{previous}
We can separate NIKA2 into four main sub-systems: 

\begin{itemize}

\item {\bfseries Optics:} the instantaneous field-of-View is 6.5~arcmin. Being M1 and M2 the primary 30~m mirror and M2 the sub-reflector, the whole optical chain consists of several mirrors at room temperature (M3, M4, M5 and M6) and cold optics inside the cryostat (see Fig.~\ref{fig:opt}). The cold optics consists  of aluminium mirrors (M7 and M8 cooled at 80~K) and cold refractive optics (high density polyethylene - HDPE lenses placed from the 1~K stage to the 100~mK stage). The rejection of out-of-band emission from the sky and the telescope is achieved by using a series of low-pass metal mesh filters, placed at different cryogenic stages in order to
minimize the thermal loading on the detectors. The band splitting between 2 and 1.15~mm channels is achieved by a dichroic installed at the 100~mK stage (see the full list of NIKA2 optical filters in Tab. 1). The polarisation facilities consist of a multi-mesh hot-pressed Half-Wave-Plate (HWP)\cite{pisano2016} mounted in front of the NIKA2 cryostat window in a mechanical modulator performing the rotation thanks to a step motor. Since the LEKID design used in the NIKA2 instrument is sensitive to both linear polarisations, the beam is further split by a wire grid polariser at 100 mK cryogenic stage in order to analyse the modulated linear polarisation into two arrays observing at 1.15~mm without any waste of signal. 

\item {\bfseries Cryostat:}
The nominal working temperature of 150~mK is achieved by two 4~K cryocooler and a closed-cycle $^3He$-$^4He$ dilution fridge. The cool-down process is remotely controlled and does not require cryogenics liquids. The whole process lasts about 5 days with four full days of pre-cooling and about one day of dilution cool-down. The fluctuation of the detectors temperature is of the order of 0.1\,mK RMS over the duration of a typical scan (15 minutes). 

\begin{figure}[t!]
\begin{center}
\includegraphics[angle=270,width=0.75\textwidth]{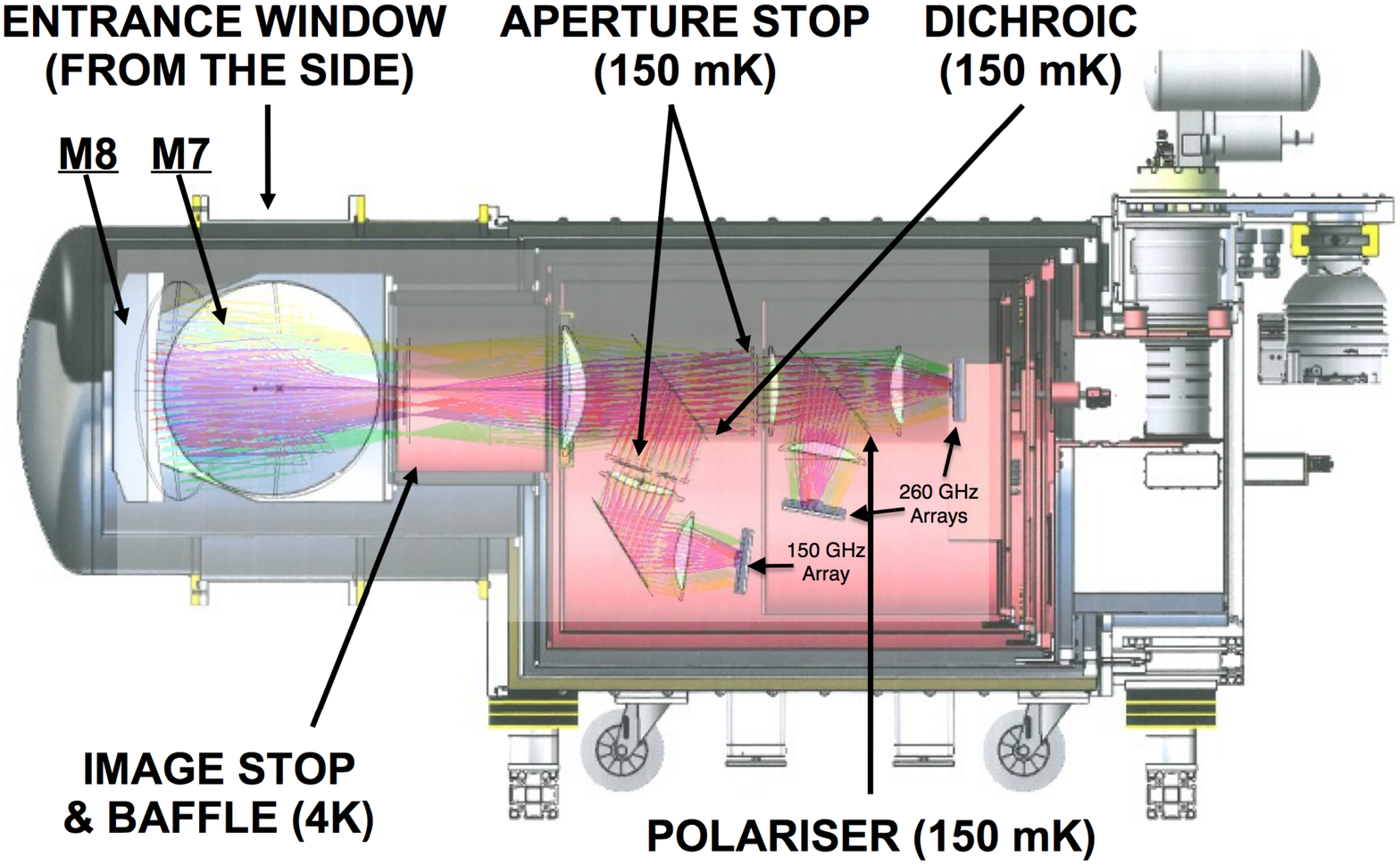}
\caption{Cross-section of the NIKA2 cryostat with the whole cold optical chain\cite{nika2}.
\label{fig:opt}}
\end{center}
\end{figure}

\begin{table}[b!]
\label{tab:fil}
\begin{center}       
\begin{tabular}{||c | c||} 
\hline
\hline
{\bfseries Common Filters} & Working Temperature [K]  \\
\hline
\hline
Thermal filter CT15  &     300  \\
Thermal filter CT15  &     150  \\
Thermal filter CT30  &     80  \\
Thermal filter CT30  &     80  \\
W1924 13cm-1 LPE  &     80  \\
W1923 12cm-1 LPE  &     4  \\
W1922 11cm-1 LPE  &     1  \\
dichroic 6.1cm-1 HPE  &     0.1  \\
W1924 13cm-1 LPE at 80K  &     80  \\
\hline 
\hline
{\bfseries Band-defining Filters} &  \\
\hline
\hline
K1931 5.65cm-1 LPE  & 0.1 (on 2mm) \\
K1670 4.0cm-1 HPE & 0.1 (on 2mm) \\
K1930 10.15cm-1 & 0.1 (on 1mm) \\
W2217 6.5cm-1 & 0.1 (on 1mm) \\
\hline 
\hline
\end{tabular}
\end{center}
\caption{List of NIKA2 optical filters.} 
\end{table}

\begin{table}[t!]
\label{tablereu}
\begin{center}       
\begin{tabular}{||c | c c||} 
\hline
\hline
{\bfseries Wavelength} [mm] & 2 & 1.15 \\
\hline
\hline
{\bfseries Average KID per feed-line}   [\#] & 255   &     142.5 \\
{\bfseries Board count}  [\#] & 4   &     16  \\
{\bfseries Total Power consumption}  [W] & 370  &   1220 \\
{\bfseries Tone tuning resolution} [Hz] & 953 &  953 \\
{\bfseries Frequency range}  [GHz] &1.3-1.8 &        1.9-2.4 \\
\hline 
\hline
\end{tabular}
\end{center}
\caption{Main characteristics of the NIKA2 readout.}\label{tablereu}
\end{table}

\item {\bfseries Detectors:}

NIKA2 Lumped Element Kinetic Inductance Detectors (LEKIDs) are composed of an interdigitated capacitor and a meander inductor acting as an absorber. The mask is designed with an Hilbert pattern to absorb both polarisations\cite{Roesch2012,goupy2}. 
Each array is fabricated on a single 4-inches high-resistivity silicon wafer (thickness equal to 250 and 300~$\mu m$ for the 1.15~mm and 2~mm arrays respectively). The pixels pattern is wet etched from a thin aluminium film (18~nm) deposited by e-beam evaporation. The use of thin aluminium film has several advantages: a better match with free space impedance of the incoming photons and a high kinetic inductance. This kind of film has been largely adopted for NIKA and in general for all the applications in our laboratory\cite{goupyLTD,catalano_3mm}. 
In order to exploit all the angular resolution of the IRAM 30~m telescope, the pixel sizes have been set to 2~mm for the 1.15~mm arrays and 2.3~mm for 2~mm array. This corresponds to a pixel size in units F$\lambda$ ( $\lambda$ = wavelength and F= f-number or relative aperture) equal to 1 for all the arrays. 
The geometrical coupling between pixels and the corresponding feed-line is made with a microstrip readout line. The consequence of this geometry is that the detectors must be front-illuminated. More details about the design and fabrication of NIKA detectors can be found in \cite{nika2,goupyLTD,Roesch2012}.

\item {\bfseries Readout electronics:}
The readout comprises coaxial cables connected from 300~K to the base temperature, 20 low-noise cryogenics amplifiers (LNA) installed in the 4~K stage, and warm electronics. The latter is constituted by readout boards named New Iram Kid ELectronic in Advanced Mezzanine Card format (NIKEL\_AMC), central, clocking and synchronisation boards (CCSB) mounted on the MicroTCA Carrier Hub (MCH) and one 600\,W power supply. These boards are distributed in three micro-Telecommunication Computing Architecture (MTCA) crates. In order to readout 2896 pixels, NIKA2 is equipped with 3 crates hosting 20 boards (8X2 for arrays at 1.15~mm and 4x1 for 2~mm array). In Tab. \ref{tablereu} are presented the main characteristics of the NIKA~2 readout. For a full description of the readout electronics please see for example\cite{Bourrion2016}. 

\end{itemize}

\section{The commissioning phase: a summary of the performance}\label{LEKID}

The NIKA camera was commissioned between September 2015 and April 2017. In that period we performed several technical observing campaigns observing point like sources, in order to assess the NIKA photometry, and extended sources to demonstrate the possibility to reconstruct angular scales up to several arc-minutes.

\begin{table*}[b]
  \centering
  \caption{Summary of the principle characteristics and performance of the NIKA2 instrument\cite{nika2}. Beam efficiency here is defined as the ratio between the main beam power and the total beam power up to a radius of 250$^{\prime \prime}$.\label{sumperf}}
  \begin{tabular}{|c|c|c|c|c|}
    \hline
	Channel & \multicolumn{3}{|c|}{260 GHz} & 150 GHz \\
            & \multicolumn{3}{|c|}{1.15 mm}     &  2 mm \\ 
    \hline
    Arrays & A1 & A3  & A1\&3 & A2 \\
    \hline
    Number of designed detectors       & 1140      &  1140    &    &    616      \\
    Number of valid detectors    &  952      &   961    &   &    553      \\ 
    \hline
    FOV diameter [arcmin]     &   6.5              &  6.5              &   6.5        &    6.5        \\
    FWHM [arcsec]             &   $11.3 \pm 0.2$   &  $11.2 \pm 0.2$  &   $11.2 \pm 0.1$           &  $17.7 \pm 0.1$ \\      
    Beam efficiency [\% ]   & $55 \pm 5$  &  $53 \pm 5$  &  $60 \pm 6$        &     $75 \pm 5$ \\
    \hline 
    rms calibration error [\%]            & 4.5  & 6.6  &   & 5 \\
    \hline
    Model absolute calibration uncertainty [\%] &  \multicolumn{4}{|c|}{5} \\
    \hline
    RMS pointing error    [arcsec]    & \multicolumn{4}{|c|}{$<3$} \\
    \hline
    NEFD [mJy.s$^{1/2}$]           &    &     & $33 \pm 2$    & $8 \pm 1$ \\
    \hline 
  \end{tabular}
\end{table*}

\begin{figure}[b!]
\begin{center}
\includegraphics[angle=0,width=1.0\textwidth]{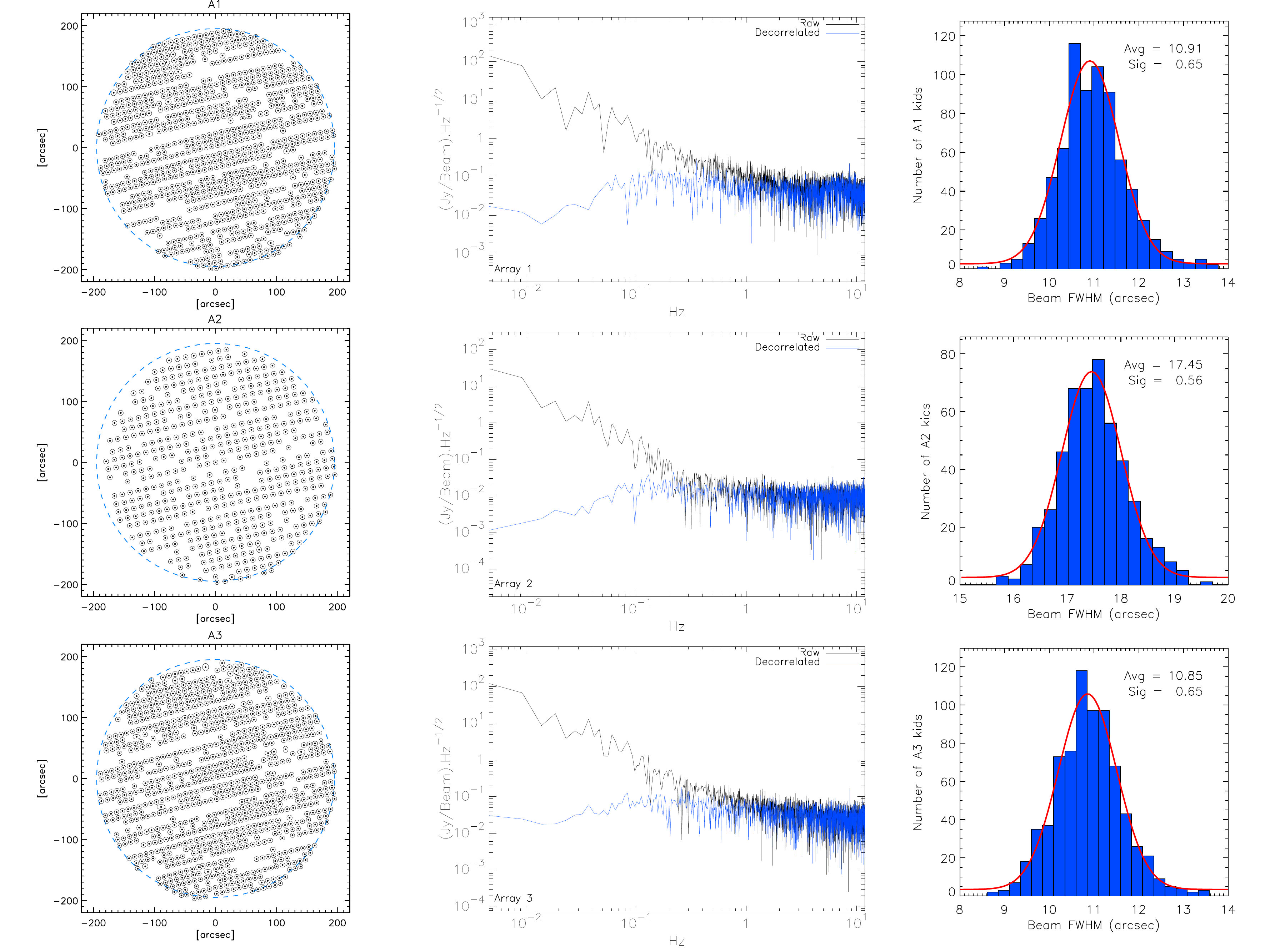}
\caption{From top to bottom left column: detectors positions for arrays A1 (260~GHz-H), A2 (150~GHz) and A3 (260~GHz-V). Central column: power spectra of the NIKA2 time ordered data before (black) and after (blue) subtraction of atmospheric fluctuations, which show-up at frequencies below 1~Hz. Right column: main beam FWHM distribution of all valid KID detectors of arrays A1, A3, and A2\cite{nika2}.
\label{fig:perf_plot}}
\end{center}
\end{figure}


Table \ref{sumperf} summarises the main NIKA characteristics and performance as measured on the sky. We obtained a sensitivity (averaged over all valid detectors) of 33 and 8~mJy$\cdot \sqrt{s}$ for the best weather conditions for the 1.15~mm (A1\&A3) arrays and for 2 mm (A2) array respectively, estimated on point-like sources. 

The sensitivity at 1.15~mm is limited by sky noise decorrelation techniques (this will be investigated in forthcoming articles Perotto et al. 2018, Ponthieu et al. 2018) and a still unidentified optical problem reducing considerably the illumination on the array A1. This issue is under investigation and will be addressed in a forthcoming publication.


In april 2017 we performed a science verification run in order to summarise the capability of NIKA 2 camera to recover large angular scales with high mapping speed. The choice of the source was on the high redshift cluster of galaxy PSZ2-G144.8 observed via the Sunyaev Zel'dovitch effect. The analysis of this cluster is still in progress and it will be published in a dedicated paper in the incoming months. 

\section{Conclusion}

NIKA2 has been successfully installed at the IRAM 30-m telescope in october 2015 as planned. The commissioning phase for the intensity observations is achieved. Acceptance meeting scheduled in early September 2017 made official the beginning of the NIKA~2 instrument as a open tool available for astronomers since summer semester 2017. 
\\
Recent results can be found on \href{}{http://lpsc.in2p3.fr/NIKA2LPSZ/nika2sz.release.php}. Finally, commissioning for polarisation shows encouraging results, the work will continue over the incoming months.

\begin{acknowledgements}
This work has been partially funded by the Foundation Nanoscience Grenoble, the LabEx FOCUS ANR-11-LABX-0013 and 
the ANR under the contracts "MKIDS", "NIKA" and ANR-15-CE31-0017. 
This work has benefited from the support of the European Research Council Advanced Grant ORISTARS 
under the European Union's Seventh Framework Programme (Grant Agreement no. 291294).
We acknowledge fundings from the ENIGMASS French LabEx (R. A. and F. R.), 
the CNES post-doctoral fellowship program (R. A.),  the CNES doctoral fellowship program (A. R.) and 
the FOCUS French LabEx doctoral fellowship program (A. R.).
\end{acknowledgements}


\begin{thebibliography}{99}
%
\bibitem[Catalano et al.(2014)]{catalano_nika2014} Catalano, A., Calvo, M., Ponthieu, N., et al.\ 2014, A\&A, 569, AA9 
%
\bibitem[Adam, R. et~al.(2014]{adam2014} Adam, R. et~al. 2014, A\&A 569, id.A66
%
\bibitem[NIKA Collaboration (2017)]{nika2} NIKA Collaboration \ 2017, arXiv:1707.00908
%
\bibitem[Pisano, G. et al. (2016)]{pisano2016} Pisano, G. et al. \ 2016 arXiv:1610.00582
%
\bibitem[Roesch et al.(2012)]{Roesch2012} Roesch, M., et al.\ 2012, arXiv:1212.4585 
%
\bibitem[Goupy et al.(2016)]{goupy2} Goupy, J., et al., JLTD, \ 2016, 184, 661-667 
%
\bibitem[Goupy et al.(2010)]{goupyLTD} Goupy, J., et al., JLTD, \ 2017, in press 
%
\bibitem[Catalano et al.(2014)]{catalano_3mm} Catalano, A. et al. \ 2015, A\&A, 580, A15 
%
\bibitem[Bourrion et al. (2016)]{Bourrion2016} Bourrion, O. et~al. \ 2016, Journal of Instrumentation 11, P11001, arXiv:1602.01288
%
%
%
%
%
%
%
%
%
%
%
%
%
%
%
%
\end{thebibliography}
\end{document}